\documentclass[aps,prl,floatfix,10pt]{revtex4-1}
\usepackage[utf8]{inputenc}
\usepackage{mathrsfs}
\usepackage{amsmath}
\usepackage{amsfonts}
\usepackage{amssymb}
\usepackage{amsthm}
\usepackage{multirow}
\usepackage{color}
\usepackage{graphicx}
\usepackage{lettrine}
\graphicspath{{./images/}}
\usepackage{geometry}
\usepackage{hyperref}
\usepackage{lineno}
\usepackage{textcomp}
\usepackage[toc,page]{appendix}
\usepackage{ulem}

\hypersetup{
     unicode=false,          
     pdftoolbar=true,        
     pdfmenubar=true,        
     pdffitwindow=false,     
     pdfstartview={FitH},    
     pdftitle={My title},    
     pdfauthor={Author},     
     pdfsubject={Subject},   
     pdfcreator={Creator},   
     pdfproducer={Producer}, 
     pdfkeywords={keyword1} {key2} {key3}, 
     pdfnewwindow=true,      
     colorlinks=false,       
     linkcolor=red,
     citecolor=green,        
     filecolor=magenta,      
     urlcolor=cyan           
}
\begin{document}
\makeatletter
\@ifundefined{textcolor}{}
{
\definecolor{BLACK}{gray}{0}
\definecolor{WHITE}{gray}{1}
\definecolor{RED}{rgb}{1,0,0}
\definecolor{GREEN}{rgb}{0,1,0}
\definecolor{BLUE}{rgb}{0,0,1}
\definecolor{CYAN}{cmyk}{1,0,0,0}
\definecolor{MAGENTA}{cmyk}{0,1,0,0}
\definecolor{YELLOW}{cmyk}{0,0,1,0}
}
\makeatother
\onecolumngrid
\begin{flushleft}
{\Large{\textbf{Square-root higher-order Weyl semimetals}}}
\quad\par
\quad\par
Lingling Song, Huanhuan Yang, Yunshan Cao, and Peng Yan$^{*}$
\quad\par
\quad\par

School of Electronic Science and Engineering and State Key Laboratory of Electronic Thin Films and Integrated Devices, University of Electronic Science and Technology of China, Chengdu 610054, China.
\end{flushleft}
\quad\par
\quad\par
{\bf \noindent The mathematical foundation of quantum mechanics is built on linear algebra, while the application of nonlinear operators can lead to outstanding discoveries under some circumstances. In this Letter, we propose a model of square-root higher-order Weyl semimetal (SHOWS) by inheriting features from its parent Hamiltonians. It is found that the SHOWS hosts both ``Fermi-arc'' surface and hinge states that connect the projection of the Weyl points. We theoretically construct and experimentally observe the exotic SHOWS state in three-dimensional (3D) stacked electric circuits with honeycomb-kagome hybridizations and double-helix interlayer couplings. Our results open the door for realizing the square-root topology in 3D solid-state platforms.}

~\\
\twocolumngrid
\noindent Nearly all the operators encountered in quantum mechanics are linear (or antilinear) operators, such as the rotation, translation, parity, time reversal, etc, which allows us to construct the mathematical basis of quantum mechanics formulated on linear algebra. Square-root operator is one of the few exceptions. Historically, Paul Dirac derived the Dirac equation through a square-root operation on the Klein-Gordon (KG) equation to describe all spin-$\frac {1} {2}$ massive particles that inherent the Lorentz-covariance of the parent KG equation \cite{Dirac1928,Greiner2003,Anderson1933}. The approach has inspired Arkinstall \emph{et al.} \cite{ArkinstallL2017} to propose the concept of square-root topological insulator (TI) by taking the nontrivial square-root of a tight-binding Hamiltonian in periodic lattices. The most appealing feature of square-root TI is that it inherits the nontrivial nature of Bloch wave function from its parent Hamiltonian. The square-root TI was subsequently observed in a photonic cage \cite{Kremer2020}. Recently, the square-root operation has been applied to higher-order topological insulators (HOTIs) that allow topologically robust edge states with codimension larger than one \cite{SongL2020,Mizoguchi2021,YanM2020,YanW2021,JungM2021,Marques2021,WuH2021,Dias2021,Yoshida2021,Marques20212}. Besides the gapped solution, e.g., the electron-positron pair, the Dirac equation allows another crucial gapless or massless solution called Weyl fermion \cite{Weyl1929} that plays an important role in quantum field theory and the Standard Model. Although not yet observed among elementary particles, Weyl fermions are shown to exist as collective excitations in Weyl semimetals \cite{Wan2011,Xu2015,Lv2015}. It is thus intriguing to ask if the square-root operation can apply to semimetals \cite{Mizoguchi20212} or higher-order semimetals \cite{WangH2020,Ghorashi2020,WeiQ2021,QiuH2021,LuoL2021}, and particularly how to realize these exotic states in experiments.

In this article, we propose a tight-binding (TB) model of the square-root higher-order Weyl semimetal (SHOWS) by a vertical stacking of two-dimensional (2D) square-root HOTIs with interlayer couplings in a double-helix  fashion. It is found that the SHOWS hosts both 2D surface arc states and one-dimensional (1D) hinge states with the topological feature being fully characterized by the quantized bulk polarization. We construct the TB model in honeycomb-kagome (HK) hybridized inductor-capacitor (\emph{LC}) circuit networks. By performing both the impedance and voltage measurements in the stacked HK circuit, we identify the fingerprint of the SHOWS by directly observing the Weyl points, the ``Fermi-arc'' surface states, and the hinge states. It is revealed that both the surface states and the hinge states ideally connect the projections of the Weyl points, consistent with theoretical calculations.\\

\begin{figure*}[htbp!]
 \centering
 \includegraphics[width=0.96\textwidth]{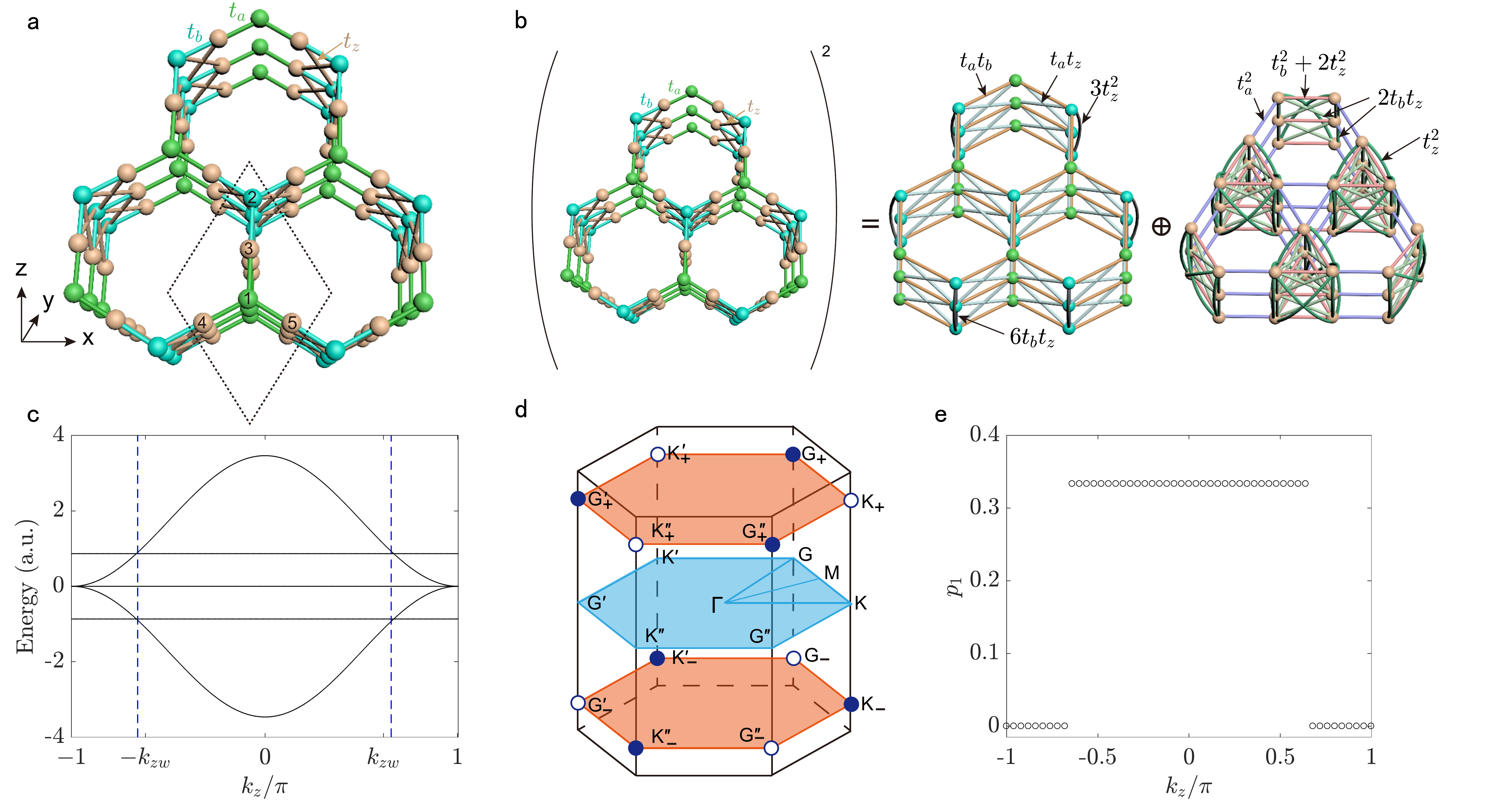}\\
 \caption{ \textbf{a} Illustration of an infinite three-dimensional (3D) stacked HK TB model. The unit cell including five nodes is represented by the dashed black rhombus. The intralayer hoppings are $t_{a}$ (green) and $t_{b}$ (blue) in the $x-y$ plane, whereas the interlayer double-helix hopping is $t_{z}$ (brown). \textbf{b} The equivalence between the squared Hamiltonian of the HK circuit and its parents. \textbf{c} The bulk dispersion along the $k_z$ direction with $(k_x,k_y)=(4\pi/3,0)$. The dashed blue line indicates the position of the degenerate points. \textbf{d} The first Brillouin zone and the distribution of the Weyl points. \textbf{e} Bulk polarization $p_1$ as a function of $k_z$. For TB calculations in \textbf{c} and \textbf{e}, we set $t_a=0.5$, $t_b=1$, and $t_z=0.5$.}\label{model_inf}
\end{figure*}

\noindent\textbf{Model}\\
Figure \ref {model_inf}\textbf{a} shows the lattice structure of the proposed model, the square of which can be viewed as the direct sum of a stacked honeycomb and a breathing kagome lattices (see Fig. \ref {model_inf}\textbf{b} and the analysis in Supplementary Information Sec. I \cite{Supplemental}). The tight-binding Hamiltonian is given by
 \begin{equation}\label{Eq1}
 \begin{aligned}
 \mathcal {H}=&t_a\sum_{\langle m,n\rangle}\left(a_m^\dagger c_n+a_m^\dagger d_n+a_m^\dagger e_n\right)\\
  &+t_b\sum_{\langle m,n\rangle}\left(b_m^\dagger c_n+b_m^\dagger d_n+b_m^\dagger e_n\right)\\
 &+t_z\sum_{\langle \langle m,n\rangle\rangle}\left(b_m^\dagger c_n+b_m^\dagger d_n+b_m^\dagger e_n\right)+{\rm H.c.},
 \end{aligned}
 \end{equation}
 where $a^\dagger$ ($a$), $b^\dagger$ ($b$), $c^\dagger$ ($c$), $d^\dagger$ ($d$), and $e^\dagger$ ($e$) are the creation (annihilation) operators on the site 1-5, respectively, $\langle m,n\rangle$ and $\langle \langle m,n\rangle\rangle$ label the nearest-neighbor and next-nearest-neighbor coupling, respectively, and $t_a$, $t_b$, and $t_z$ are the hopping parameters. Without loss of generality, we assume all hopping paramaters are positive.
 In momentum space, the Hamiltonian can be expressed as
\begin{equation}\label{Eq5}
 \mathcal {H}=\left(
 \begin{matrix}
   O_{2,2} & \Phi_{\bf k}^{\dagger}\\
  \Phi_{\bf k} & O_{3,3}\\
 \end{matrix}
 \right),
 \end{equation}
where $O_{2,2}$ and $O_{3,3}$ are the $2\times2$ and $3\times3$ zero matrix, respectively, and $\Phi_{\bf k}$ is the $3\times2$ matrix
\begin{equation}\label{Eq6}
 \Phi_{\bf k}=\left(
 \begin{matrix}
  t_{a} & t_{b}+2t_{z}\text{cos}(\mathbf{k}\cdot{\mathbf{a}_3}) \\
  t_{a} & [t_{b}+2t_{z}\text{cos}(\mathbf{k}\cdot{\mathbf{a}_3})]e^{-i\mathbf{k}\cdot{\mathbf{a}_1}} \\
  t_{a} & [t_{b}+2t_{z}\text{cos}(\mathbf{k}\cdot{\mathbf{a}_3})]e^{-i\mathbf{k}\cdot{\mathbf{a}_2}}\\
 \end{matrix}
 \right).
 \end{equation}
Here $\mathbf{k}=(k_x,k_y,k_z)$ is the wave vector, and $ \textbf {a}_{1}=\frac {1} {2} \hat {x} + \frac {\sqrt {3}} {2} \hat {y} $, $\textbf { a}_{2}=-\frac {1} {2} \hat {x} + \frac {\sqrt {3}} {2} \hat {y}$ and $\textbf {a}_{3}= \hat {z}$ are three basic vectors.

By taking the square of the original Hamiltonian \eqref{Eq5}, we can conveniently obtain the dispersion relation of $[\mathcal {H}]^{2}$ (see Supplementary Information Sec. I \cite{Supplemental})
\begin{equation}\label{Eq11}
E_{\bf k}=0\ \ \text{and}\ \frac{3}{2}\left[t_{a}^{2}+t_{b}'^{2}\pm
 \sqrt{(t_{a}^2-t_{b}'^2)^2+4t_{a}^{2}t_{b}'^{2}|\Delta({\bf k})|^{2}}\right],
\end{equation}
with
 $ t_{b}'=t_{b}+2t_z\text{cos}(k_z)$ and $\Delta({\bf k})=(1 +e^{i\mathbf{k}\cdot{\mathbf{a}_1}} + e^{i\mathbf{k}\cdot{\mathbf{a}_2}})/3 $. The band structure of the original Hamiltonian is thus given by $\varepsilon_{\bf k}=\pm\sqrt{E_{\bf k}}$. It is found that the band structure closes at the twofold degenerate points $K_\pm = (4\pi/3,0,\pm k_{zw})$, as shown in Fig. \ref{model_inf}c, with $k_{zw}=\text{arccos}[(t_a-t_b)/(2t_z)]$ when $|t_a-t_b|<2t_z$. It is straightforward to verify that their time-reversal counterparts are $G^{'}_\pm= (-4\pi/3,0,\pm k_{zw})$, and their equivalence points locate at $G_\pm$, $G^{''}_\pm$, $K^{'}_\pm$, and $K^{''}_\pm$,\begin{figure*}[htbp!]
 \centering
 \includegraphics[width=0.96\textwidth]{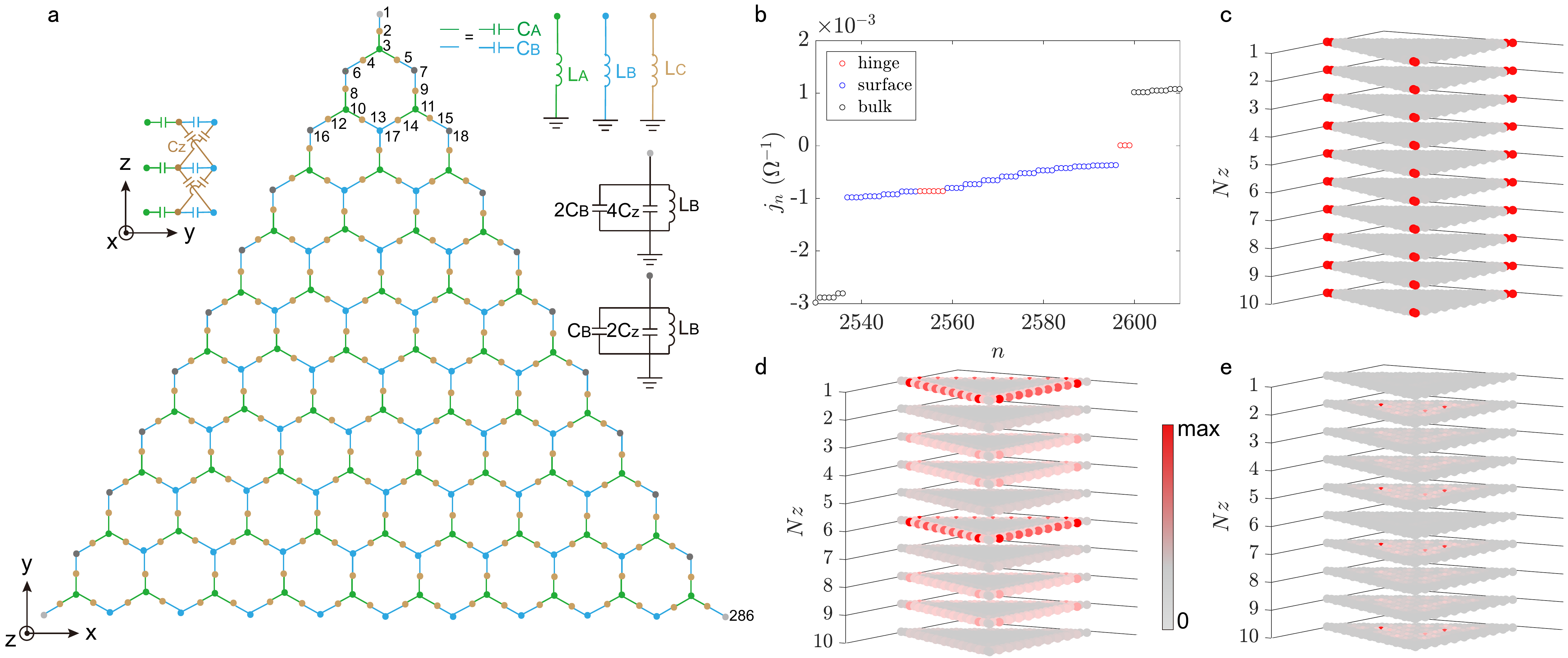}\\
 \caption{\textbf{a} Top view of a stacked 10-layer HK circuit with 2860 nodes. \textbf{b} Admittances for $C_A=C_B/2=0.5$ nF, $L_A=30$ $\mu$H, $L_B=7.5$ $\mu$H, $L_C=18$ $\mu$H, and $L_G=21.829$ $\mu$H. The red, blue, black symbols represent the hinge, surface, bulk states, respectively. Spatial distribution of the wave-function of the normalized hinge ($j_n=3.534\times10^{-7}$ $\Omega^{-1}$) state \textbf{c}, surface ($j_n=-0.0003762$ $\Omega^{-1}$) state \textbf{d}, and bulk ($j_n=0.001042$ $\Omega^{-1}$) state \textbf{e}.}\label{Energy_spectrum}
\end{figure*}
 as shown in Fig. \ref {model_inf}\textbf{d}. By evaluating the topological charge $C_{\rm FS}$, we find that the hollow and solid circles plotted in Fig. \ref {model_inf}\textbf{d} denote the Weyl points with opposite topological charges, i.e., $+1$ and $-1$, respectively (see Supplementary Information Sec. II and Fig. S1 \cite{Supplemental}). In addition, we derive the low-energy effective Hamiltonian near the degeneracy points, and obtain a linear crossing in the vicinity of the Weyl points (see Supplementary Information Sec. II \cite{Supplemental}). The computation of Berry curvatures are plotted in Figs. S1c and S1d, which indeed demonstrates that the Weyl points manifest as singularities (source and drain), a close analog to the magnetic monopole in momentum space.

For a system with the rotational symmetry (it is $C_3$ in our model), the bulk polarization is the appropriate invariant to characterize the topological features. For the $n$th band, the bulk polarization as a function of $k_z$ is written as
\begin{equation}\label{Eq14}
 2\pi p_{n}(k_z)=\text{arg} \theta_{n}(\mathbf{k})\ (\text{mod}~ 2\pi),
\end{equation}where $\mathbf{k}=(4\pi/3,0,\pm k_z)$, and $\theta_{n}({\bf k})=u^{\dag}_{n}({\bf k})U_{{\bf k}}u_{n}({\bf k})$ with $u_{n}({\bf k})$ the $n$th eigenvector and the $U$-matrix
\begin{equation}
U_{\bf k}=\left(
 \begin{matrix}
   1 & 0& 0& 0& 0\\
  0& e^{-i\mathbf{k}\cdot{\mathbf{a}_2}}& 0& 0& 0\\
  0& 0& 0& 0& 1\\
  0& 0& 1& 0& 0\\
  0& 0& 0& 1& 0\\
 \end{matrix}
 \right).
 \end{equation}
 Here we are particularly interested in the 1st (or 5th) band, because the Weyl points only appear in the intersecting between the first and second energy bands (or between the fourth and fifth energy bands). As shown in Fig. \ref{model_inf}\textbf{e}, $p_1$ takes $1/3$ for $|k_z|<|k_{zw}|$, and 0 for $|k_z|>|k_{zw}|$. The topological phase transition occurs at $k_z=\pm k_{zw}$. A non-vanishing $p_1$ indicates the very presence of the higher-order topologial edge states. The present model unambiguously demonstrates the bulk-hinge correspondence and manifests itself as an ideal SHOWS (see Fig. S2\textbf{c} in Supplementary Information Sec. III \cite{Supplemental}). It is noted that a pair of Weyl points emerge with opposite wave vectors (see Fig. \ref {model_inf}\textbf{c}) because of the inversion-symmetry breaking in our model. It is worth mentioning that the present model also allows a 3D square-root HOTI phase (see Figs. S2\textbf{d}-S2\textbf{f} in Supplementary Information Sec. III \cite{Supplemental}). In what follows, we construct the tight-binding SHOWS model in 3D stacked HK \emph{LC} circuits.\\
\noindent{\bf Circuit realization of SHOWS}\\
We consider a stacked 10-layer HK circuit with $\mathcal{N} = 2860$ nodes, as depicted in Fig. \ref{Energy_spectrum}\textbf{a}.
The circuit dynamics at frequency $\omega$ obeys Kirchhoff's law
$I_a(\omega)=\sum_bJ_{ab}(\omega)V_b(\omega)$, with $I_a$ the external current flowing into node $a$, $V_b$ the voltage of node $b$, and $J_{ab}$($\omega$) being the circuit Laplacian
\begin{equation}\label{Eq7}
J(\omega)=\left(
\begin{array}{ccccccc}
J_{0B} & -J_B & 0  & 0  &  0  &  0 & \ldots\\
-J_B & J_{0C}  & -J_A & 0  &  0  &  0 & \ldots \\
0 & -J_A & J_{0A} & -J_A  & -J_A &  0 & \ldots\\
0  & 0 & -J_A  & J_{0C} & 0  & -J_B & \ldots\\
0  & 0   & -J_A & 0  & J_{0C} &  0 & \ldots\\
0  & 0   & 0 & -J_B  & 0 &  J_{0B}& \ldots\\
\vdots & \vdots & \vdots & \vdots  & \vdots  & \vdots& \ddots \\
\end{array}
\right)_{\mathcal{N}\times \mathcal{N}},
\end{equation}
with $J_{0A}=3i\omega C_{A}+1/(i\omega L_{A})$, $J_{0B}=3i\omega C_{B}+1/(i\omega L_B)$, $J_{0C}=i\omega (C_A+C_B)+1/(i\omega L_C)$, $J_{A}=i\omega C_{A}$, and $J_{B}=i\omega (C_{B}+2C_Z)$. Under the resonance condition $\omega_0=1/\sqrt{3C_AL_A}=1/\sqrt{(3C_B+6C_Z)L_B}$=$1/\sqrt{(C_A+C_B+2C_Z)L_C}$, the circuit Laplacian \eqref{Eq7} exactly recovers the tight-binding Hamiltonian by the following one-to-one correspondence: $-\omega_0C_A\leftrightarrow{t_a}$, $-\omega_0C_B\leftrightarrow{t_b}$, and $-\omega_0C_Z\leftrightarrow{t_z}$. To explore the square-root topological semimetal phase, we set $C_A=C_B/2=0.5$ nF, $C_Z=0.5$ nF and $L_A=30$ $\mu$H, $L_B=7.5$ $\mu$H, and $L_C=18$ $\mu$H in the following calculations, if not stated otherwise.

\begin{figure*}[htbp!]
 \centering
 \includegraphics[width=0.96\textwidth]{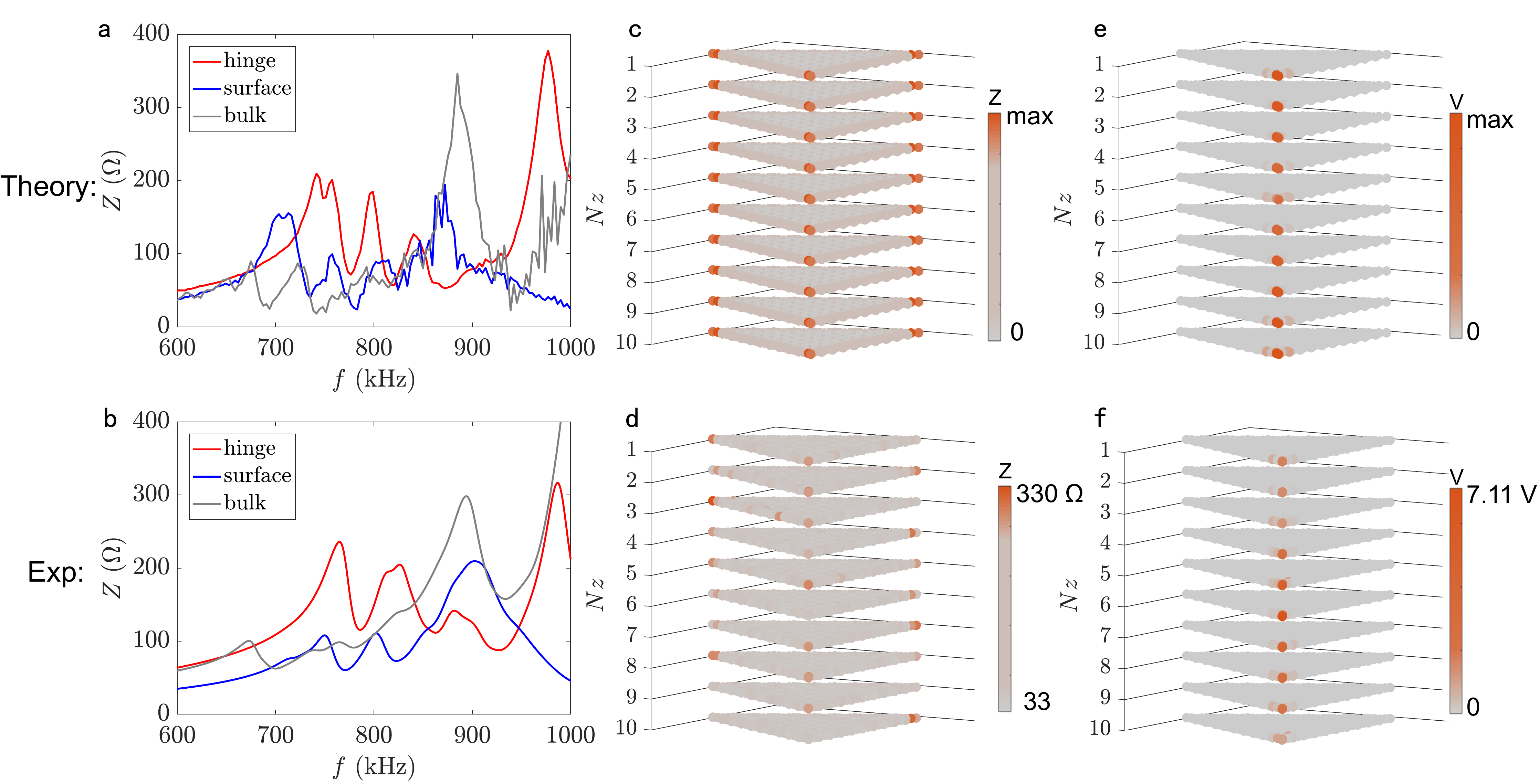}\\
 \caption{\textbf{a} Theoretical impedance versus the driving frequency in a disordered circuit. \textbf{b} Measured impedance as a function of the frequency. Calculated \textbf{c} and measured \textbf{d} impedance distribution of hinge state over the system. Hinge-state voltage distribution in theory \textbf{e} and experiment \textbf{f}.}
 \label{f6zzzz}
\end{figure*}

To facilitate the detection of the hinge states through a direct two-point impedance measurement \cite{Yang2020}, we connect a grounded inductor $L_G=22$ $\mu$H to all nodes to move the hinge modes to the zero admittance without modifying their wave functions \cite{SongL2020}. By measuring the impedance, one can precisely characterize the wave function of the zero-energy hinge states in the circuit \cite{Yang2020,SongL2020}. Figure \ref{Energy_spectrum}\textbf{b} exhibits the corresponding admittance spectrum, where the red, blue, and black dots represent the hinge, surface, and bulk states, respectively. It is obvious to see the three-fold degeneracy of the in-gap hinge states. The spatial distributions of each mode are plotted in Figs. \ref {Energy_spectrum}\textbf{c}-\ref {Energy_spectrum}\textbf{e}, from which one can straightforwardly distinguish them.

\begin{figure}[t!]
 \centering
 \includegraphics[width=0.48\textwidth]{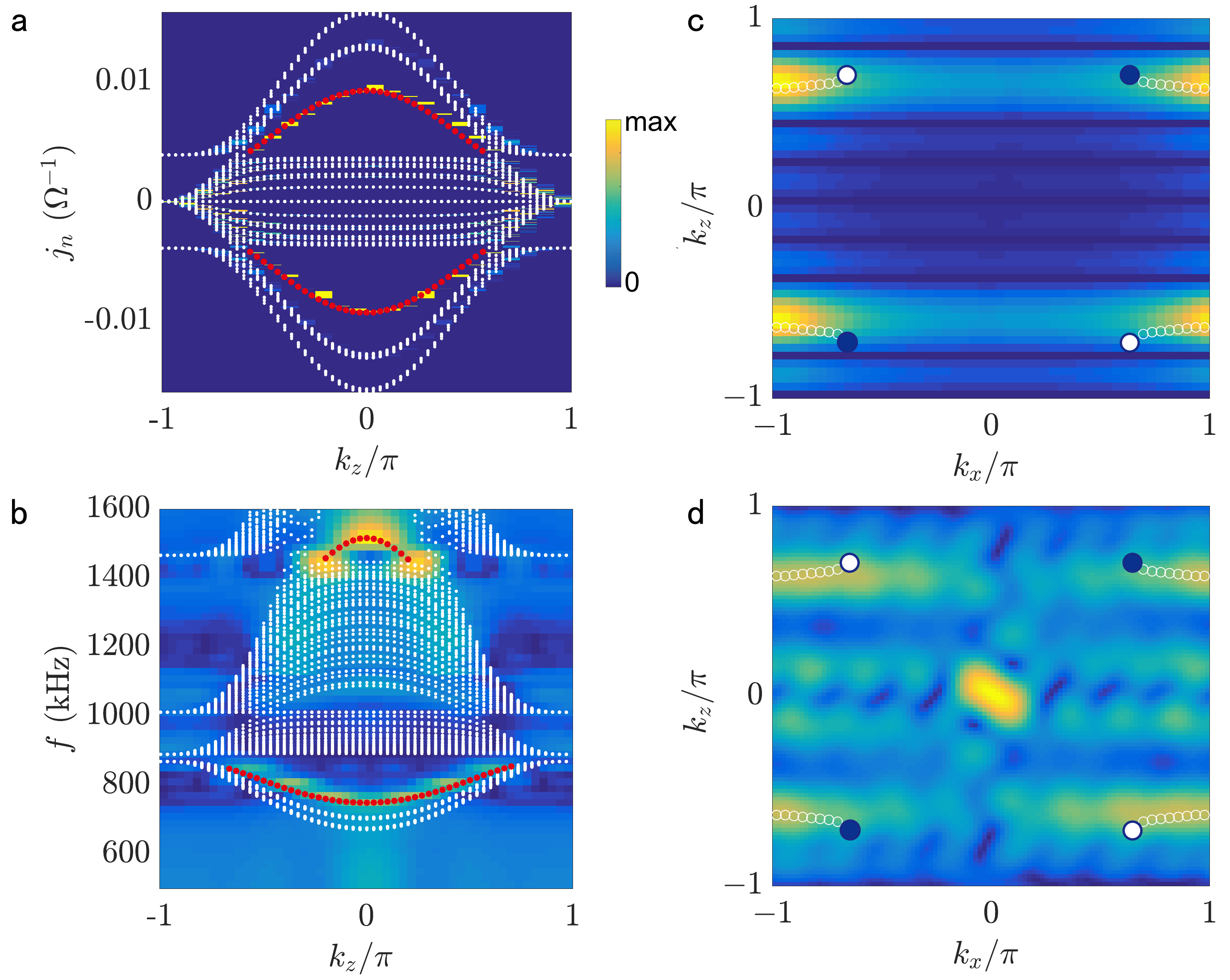}\\
 \caption{\textbf{a} The projected admittances along the $k_z$ direction. \textbf{b} Hinge state dispersion. The red dots and the colour map in \textbf{a} and \textbf{b} represent the theory and experimental hinge spectrum, respectively. \textbf{c} The numerical ``Fermi arc'' of the surface states at $j_n=0.004082$ $\Omega^{-1}$, which connects the projections of the Weyl points (the hollow and solid dots). \textbf{d} ``Fermi arc'' of the surface states at 860 kHz. The colour map and the white circles represent the experimental and theoretical results, respectively.}
 \label{f4FKZ}
\end{figure}
 The photograph of 3D \emph{LC} electric circuits fabricated on a printed circuit board is displayed in Fig. 5 in Methods. We choose electric elements $C_A=C_B/2=0.5$ nF, $C_Z=0.5$ nF and $L_A=33$ $\mu$H, $L_B=7.5$ $\mu$H, $L_C=18$ $\mu$H and $L_G=22$ $\mu$H, the same as those for theoretical computations above, but with a practical $2\%$ tolerance. The resonant frequency is then $f_c=1/(2\pi\sqrt{3C_AL_A})=755$ kHz. We first measure the impedance between three representative nodes and the ground as a function of the exciting frequency with the impedance analyzer (Keysight E4990A). Experimental results are shown in Fig. \ref{f6zzzz}\textbf{b}, which well agree with theoretical calculations plotted in Fig. \ref{f6zzzz}\textbf{a}. Here we select the 1432th, 1564th, and 1711th nodes to characterize the properties of the hinge, surface, and bulk states, respectively. We then measure the spatial distribution of the impedance and voltage over the circuit (see Figs. \ref{f6zzzz}\textbf{d} and \ref{f6zzzz}\textbf{f}), which compare reasonably well with the theoretical results plotted in Figs. \ref{f6zzzz}\textbf{c} and \ref{f6zzzz}\textbf{e}.

To characterize the hinge states more carefully, we project the dispersion to the $k_z$ axis, as shown by the color map in Fig. \ref{f4FKZ}\textbf{a}. In numerical simulations, one can conveniently take different $j_n$ and analyze the spectrum subsequently, but we cannot set the specific value of $j_n$ in circuit experiments. Fortunately, by mapping Kirchhoff's law to the Schr\"{o}dinger equation in circuit \cite{Hofmann2019,Yang2021}, we obtain the frequency dispersion (see Supplementary Information Sec. V \cite{Supplemental}) that significantly facilitates our experimental measurements. Experimentally, we impose a voltage source in the middle of one hinge of the circuits, and scan the voltage distribution along the hinge. Specifically, we input a signal
$v_s(t)$ = 5sin$(\omega t)$ V at a hinge node with the arbitrary function generator (GW AFG-3022), and then collect the voltage $v(\omega, z)$ with frequency $f=\omega/(2\pi)$ ranging from 500 kHz to 1600 kHz by using the oscilloscope (Keysight MSOX3024A). We perform the Fourier transformation on the $v(\omega, z)$ and obtain the projected dispersion along the $k_z$ direction, shown by the color map in Fig. \ref{f4FKZ}\textbf{b}. It can be seen that the hinge states connecting two Weyl points at a resonant frequency around 755 kHz, which perfectly agrees with the simulation results marked by the solid red circles.

Furthermore, it is known that the ``Fermi arc'' surface state is an unique feature of Weyl semimetals. Figure \ref{f4FKZ}\textbf{c} shows the ``Fermi arc'' surface dispersion at $j_n=0.004082$ $\Omega^{-1}$. Figure \ref{f4FKZ}\textbf{d} shows the ``Fermi arc'' surface dispersion at $f=860$ kHz. The colour map represents the measured data and the white circles denote the simulated equal-admittance contour, whereas the hollow and solid dots denote the projections of Weyl points with opposite topological charges $+1$ and $-1$, respectively. Our experiment therefore unambiguously supports the bulk-hinge correspondence and identifies the emergence of SHOWS.
~\\
\noindent{\bf Conclusions}\\
To summarize, we proposed a TB model of the SHOWS and constructed it in 3D double-helix stacked \emph{LC} circuits. Through the impedance and voltage measurements, we directly observed both the 1D prismatic states and the 2D ``Fermi arc'' surface states connecting the projected Weyl points, the fingerprint of SHOWS. Comparing with the normal Weyl semimetal \cite{Wan2011}, the SHOWS supports robust hinge states, besides the arc surface states. The emergence of Weyl pairs in SHOWS with both positive and negative energies marks its difference from the conventional higher-order Weyl semimetals \cite{WangH2020,WeiQ2021,LuoL2021}. One of the parent sublattices, i.e.,  the honeycomb lattice, originally does not support any hinge states or flat-band states. The square-root operator, however, makes it inherit these exotic states from the other parent sublattice. Our results pave the way to realizing the square-root higher-order topological states, and may inspire the exploration in other solid-state systems, such as cold atoms, photonic crystals, and elastic lattices.

~\\
\noindent{\bf Methods}\\
{\small
\noindent{\bf PCB image in experiments and circuit Laplacians.}

\begin{figure}[htbp!]
 \centering
 \includegraphics[width=0.48\textwidth]{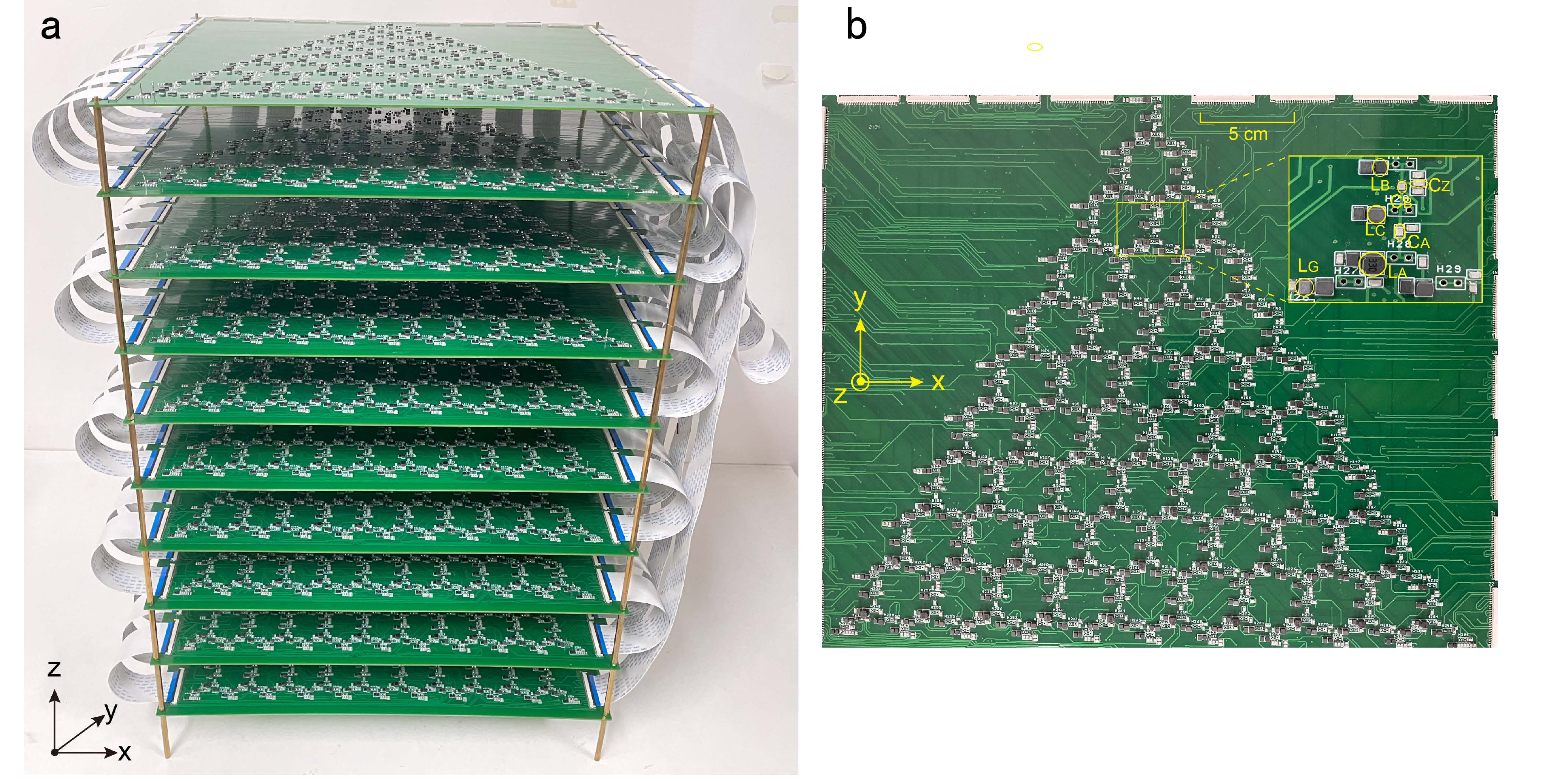}\\
 \caption{Side \textbf{a} and top \textbf{b} view of the printed circuit board in experiment. }\label{pcb}
\end{figure}

The circuit dynamics at frequency $\omega$ obeys Kirchhoff's law
$I_a(\omega)=\sum_bJ_{ab}(\omega)V_b(\omega)$, with $I_a$ the external current flowing into node $a$, $V_b$ the voltage of node $b$, and $J_{ab}$($\omega$) being the circuit Laplacian
\begin{equation}
J_{ab}(\omega)=i\mathcal {H}_{ab}(\omega)=i\omega\left[-C_{ab}+\delta_{ab}\left(\sum_nC_{an}-\frac{1}{\omega^2L_a}\right)\right],
\end{equation}
where $C_{ab}$ is the capacitance between $a$ and $b$ nodes, $L_a$ is the grounding inductance of node $a$, and the sum is taken over all nearest-neighboring nodes. For a finite circuit, the Laplacian of the circuit can be written as Eq. (7) in the main text. Considering the resonance condition $\omega=\omega_0$, one can obtain all eigenvalues $j_n$ (admittances) and eigenfunctions $\psi_n$ ($n=1,2,...,\mathcal{N}$). We set $C_A=C_B/2=0.5$ nF, $C_Z=0.5$ nF, $L_A=30$ $\mu$H, $L_B=7.5$ $\mu$H, and $L_C=18$ $\mu$H in the following calculations. In the calculation of Fig. 2 in the main text, we consider the ideal situation that all inductors and capacitors have no loss and disorder. Considering the practical loss and tolerance of capacitors and inductors, we introduced $2\%$ disorder to each capacitor and inductor in theoretical calculations thereafter. In experiments, we stacked 10 identical 2D printed circuit boards (PCBs) along the $z$ direction, as shown in Fig. \ref{pcb}\textbf{a}. Figure \ref{pcb}\textbf{b} shows the top view of the PCB with the inset zooming in the design details of the electrical circuit.
}

~\\
\noindent\textbf{Data availability}\\
The data that support the plots within this paper and other findings of
this study are available from the corresponding author on reasonable request.\\

\noindent{\large

\quad\par
\noindent{\large\textbf{Acknowledgement}}\\
We acknowledge Z.-X. Li for useful discussions. This work was supported by the National Natural Science Foundation of China (Grants No. 12074057, No. 11704060, and No. 11604041).\\

\noindent{\large\textbf{Author contributions}}\\
P.Y. conceived the idea and contributed to the project design. L.S. and H.Y. designed the circuits and performed the measurements.  L.S. developed the theory and wrote the manuscript. All authors discussed the results and revised the manuscript.\\

\noindent{\large\textbf{Competing interests}}\\
The authors declare no competing interests.\\

\noindent{\textbf{Correspondence and requests for materials} should be addressed
to P.Y. by yan@uestc.edu.cn}\\

\clearpage

\end{document}


\makeatletter
\@ifundefined{textcolor}{}
{
\definecolor{BLACK}{gray}{0}
\definecolor{WHITE}{gray}{1}
\definecolor{RED}{rgb}{1,0,0}
\definecolor{GREEN}{rgb}{0,1,0}
\definecolor{BLUE}{rgb}{0,0,1}
\definecolor{CYAN}{cmyk}{1,0,0,0}
\definecolor{MAGENTA}{cmyk}{0,1,0,0}
\definecolor{YELLOW}{cmyk}{0,0,1,0}
}
\makeatother
\onecolumngrid
\def\ooint{{\bigcirc}\kern-11.5pt{\int}\kern-6.5pt{\int}}
\begin{flushleft}
\center{\LARGE{\textbf{Supplementary Information}}}
\\[0.5cm]

{\Large{\textbf{Square-root higher-order Weyl semimetals}}}
\quad\par
\quad\par
Lingling Song, Huanhuan Yang, Yunshan Cao, and Peng Yan
\quad\par
\quad\par

School of Electronic Science and Engineering and State Key Laboratory of Electronic Thin Films and Integrated Devices, University of Electronic Science and Technology of China, Chengdu 610054, China.
\end{flushleft}

\section{ I. The squared Hamiltonian}

 It is noted that $\mathcal {H}$ [Eq. (2) in the main text] is chiral-symmetric, because it meets the condition $\mathcal {H} = -\gamma \mathcal{H}\gamma$ with
\begin{equation}\label{Eq1}
  \gamma=\left(
 \begin{matrix}
     I_{2,2} & O_{2,3}\\
    O_{3,2} & -I_{3,3}\\
 \end{matrix}
 \right),
\end{equation}
where $O_{2,3}$ $(I_{2,2})$  and $O_{3,2}$ $(I_{3,3})$ are the $2\times3$ $(2\times2)$ and $3\times2$ $(3\times3)$  zero (identity) matrixs, respectively. The Hamiltonian $\mathcal {H}$ with chiral symmetry indicates the existence of a parent Hamiltonian $\mathcal {H}^2$. With the help of parent Hamiltonian, one can obtain the eigenvalues of $\mathcal {H}$ by taking its square
\begin{equation}\label{Eq2}
\mathcal {H}^{2}=\left(
 \begin{matrix}
     h_{\bf k}^{H} & O_{2,3}\\
    O_{3,2} & h_{\bf k}^{K}\\
 \end{matrix}
 \right),
\end{equation}
where $h_{\bf k}^{H}=\Phi_{\bf k}^{\dag}\Phi_{\bf k}$ and $h_{\bf k}^{K}=\Phi_{\bf k}\Phi_{\bf k}^{\dag}$ represent the Hamiltonian of a stacked honeycomb sublattice and breathing kagome sublattice with on-site potentials, respectively. Their explicit expressions are
\begin{equation}
h_{\bf k}^{H}=\left(
  \begin{array}{cc}
    h_{11} & h_{12}  \\
    h_{12}^{*} & h_{22} \\
  \end{array}
\right),
\end{equation}
with \begin{equation}
\begin{aligned}
h_{11}&=3t_a^2,\\
h_{12}&=t_at_b+2t_at_z\cos k_z+(t_at_b+2t_at_z\cos k_z)e^{-i{\bf k}\cdot{\bf a}_1}+(t_at_b+2t_at_z\cos k_z)e^{-i{\bf k}\cdot{\bf a}_2},\\
h_{22}&=3t_b^2+12t_bt_z\cos k_z+6t_z^2+6t_z^2\cos (2k_z) ,\\
\end{aligned}
\end{equation}

 and
\begin{equation}\label{Eqs5}
h_{\bf k}^{K}=\left(
  \begin{array}{ccc}
h_{33} & h_{34} & h_{35} \\
h_{34}^{*} & h_{44} & h_{45} \\
h_{35}^{*} & h_{45}^{*} & h_{55} \\
  \end{array}
\right),
\end{equation}
with \begin{equation}
\begin{aligned}
h_{33}&=h_{44}=h_{55}=t_a^2+t_b^2+2t_z^2+4t_bt_z\cos k_z+2t_z^2\cos (2k_z) ,\\
h_{34}&=t_a^2+[t_b^2+2t_z^2+4t_bt_z\cos k_z+2t_z^2\cos (2k_z)]e^{i{\bf k}\cdot{\bf a}_1},\\
h_{35}&=t_a^2+[t_b^2+2t_z^2+4t_bt_z\cos k_z+2t_z^2\cos (2k_z)]e^{i{\bf k}\cdot{\bf a}_2}, \\
h_{45}&=t_a^2+[t_b^2+2t_z^2+4t_bt_z\cos k_z+2t_z^2\cos (2k_z)]e^{-i{\bf k}\cdot({\bf a}_1-{\bf a}_2)}.\\
\end{aligned}
\end{equation}
 We note that $h_{\bf k}^{H}$ and $h_{\bf k}^{K}$ have the same energy band solution, except that $h_{\bf k}^{K}$ has an additional flat band pinned to zero energy. The energy band solution of the $h_{\bf k}^{K}$ is
\begin{equation}\label{Eq11}
E_{\bf k}=0\ \ \text{and}\ \frac{3}{2}\left[t_{a}^{2}+t_{b}'^{2}\pm
 \sqrt{(t_{a}^2-t_{b}'^2)^2+4t_{a}^{2}t_{b}'^{2}|\Delta({\bf k})|^{2}}\right],
\end{equation}
with
 $ t_{b}'=t_{b}+2t_z\text{cos}(k_z)$ and $\Delta({\bf k})=(1 +e^{i\mathbf{k}\cdot{\mathbf{a}_1}} + e^{i\mathbf{k}\cdot{\mathbf{a}_2}})/3 $. The band structure of the original Hamiltonian is therefore given by $\varepsilon_{\bf k}=\pm\sqrt{E_{\bf k}}$.
\section{II. The linear admittance spectrum near the Weyl point and the Berry curvature}
\begin{figure*}[h!]
  \centering
  \includegraphics[width=0.8\textwidth]{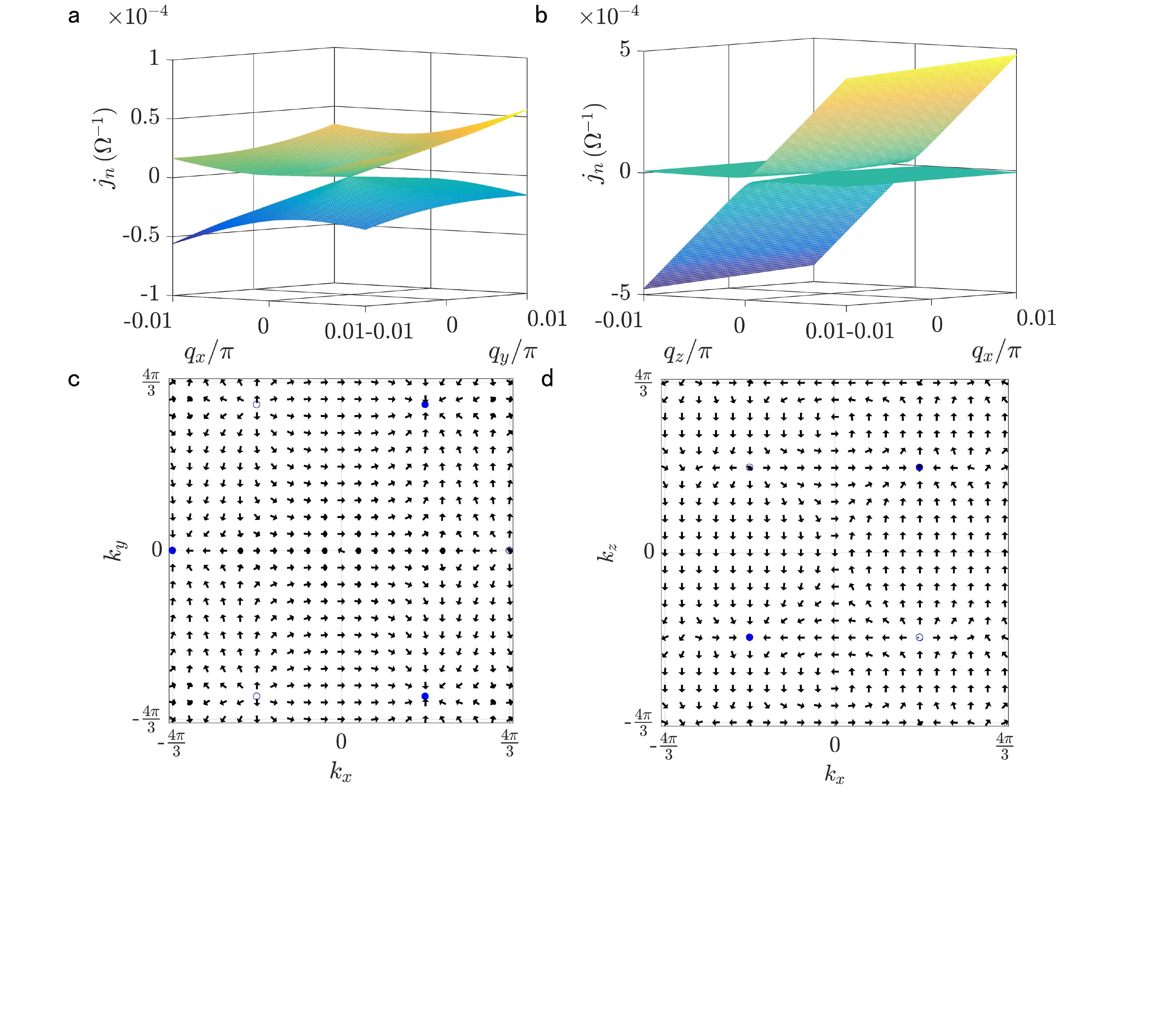}\\
  \caption{The admittance dispersion around the Weyl point $K_+$ in the \textbf{a} $q_x-q_y$ and \textbf{b} $q_x-q_z$ planes. The spatial distribution of the Berry curvature around \textbf{c} $K_+$
and \textbf{d} $K_-$. Open and solid circles represent the Weyl points with opposite topological charges $+1$ and $-1$. }\label{FS1BC}
\end{figure*}
In this section, we demonstrate that the Weyl semimetal in our system hosts linear dispersion in all three dimensions in the vicinity of the Weyl points which act like monopoles of Berry curvature. To this end, we expand $h_{\bf k}^{H}$ in terms of Pauli matrix $h_{\bf k}^{H}=\lambda_0\sigma_0+\lambda_x\sigma_x+\lambda_y\sigma_y+\lambda_z\sigma_z$  with $\sigma_{0}$ the identity matrix, $\sigma_{x}$, $\sigma_{y}$ and $\sigma_{z}$ being the Pauli matrices. The parameters $\lambda_{i}$ ($i=0,x,y,z$) are explicitly expressed as
\begin{equation}
\begin{aligned}
\lambda_{0}&=\frac{3}{2}t_a^2+\frac{3}{2}t_b^2+6t_bt_z\cos k_z+3t_z^2+3t_z^2\cos(2k_z),\\
\lambda_{x}&=t_at_b+2t_at_z\cos k_z+2(t_at_b+2t_at_z\cos k_z)\cos(\frac{1}{2}k_x)\cos(\frac{\sqrt{3}}{2}k_y),\\
\lambda_{y}&=2t_at_b\cos(\frac{1}{2}k_x)\sin(\frac{\sqrt{3}}{2}k_y)+4t_at_z\cos(\frac{1}{2}k_x)\sin(\frac{\sqrt{3}}{2}k_y)\cos k_z,\\
\lambda_{z}&=\frac{3}{2}t_a^2-\frac{3}{2}t_b^2-6t_bt_z\cos k_z-3t_z^2-3t_z^2\cos(2k_z).\\
\end{aligned}
\end{equation}
Near the point ${\bf K_+}= (4\pi/3,0,k_{zw})$, using the Taylor expansion, the parameters $\lambda_{i}$ ($i=0,x,y,z$) of the effective Hamiltonian can written as:
\begin{equation}\label{Eq9}
\begin{aligned}
\lambda_{0}&=\frac{3}{2}t_a^2+\frac{3}{2}t_b^2+\frac{3}{2}t_z^2+\frac{3}{2}t_bt_z-3\sqrt{3}(t_bt_z-t_z^2){ q_z},\\
\lambda_{x}&=t_at_z-\frac{\sqrt{3}}{2}(t_at_b-t_at_z){q_x}-\sqrt{3}t_at_z{q_z},\\
\lambda_{y}&=-\frac{\sqrt{3}}{2}(t_at_b-t_at_z){q_y},\\
\lambda_{z}&=\frac{3}{2}t_a^2+\frac{3}{2}t_b^2-\frac{3}{2}t_z^2-\frac{3}{2}t_bt_z+3\sqrt{3}(t_bt_z-t_z^2){q_z},\\
\end{aligned}
\end{equation}
with $\bf{q}=\bf{k}-\bf{K_+}$. From Eqs. \eqref{Eq9}, one can clearly see that the band linearly touches at $K_+$, which is a typical feature of band crossing of Weyl semimetals. The energy bands around the Weyl point of SHOWS inherited from \eqref{Eq9} are also linear. We then investigate the distribution of Berry curvature in momentum based on the low-energy effective Hamiltonian expanding around the Weyl points. It will be demonstrated that the Weyl points will generate Fermi arc states on the surface.
We first consider the degenerate point at $K_+$. Here, we plot the 3D band dispersion around $K_+$ in Figs. \ref{FS1BC}\textbf{a} and \ref{FS1BC}\textbf{b}. The band dispersion around $K_-$ is similar to the case around $K_+$. Obviously, the band dispersion around the degenerate points along any direction is linear. Furthermore, the Berry curvature is expressed as
 \begin{equation}\label{Eq14}
{ F_x=\frac{\partial{A_z}}{\partial{q_y}}-\frac{\partial{A_y}}{\partial{q_z}}},\\
{ F_y=\frac{\partial{A_x}}{\partial{q_z}}-\frac{\partial{A_z}}{\partial{q_x}}},\\
{ F_z=\frac{\partial{A_y}}{\partial{q_x}}-\frac{\partial{A_x}}{\partial{q_y}}},\\
\end{equation}
 where $A_\mu=-i\langle\phi|\nabla_\mu|\phi\rangle$ is the berry connection, with $\mu=x,y,z$ and $\phi({\bf q})$ being its wave function. Figures \ref{FS1BC} \textbf{c} and \ref{FS1BC}\textbf{d} show that the flux of the Berry curvature flowing from $K_+$ to $K_-$, which is similar to the magnetic monopole in momentum space.
 The monopole charge is defined as
  \begin{equation}\label{Eqcfs}
 C_{\rm FS}=\frac{1}{2\pi}\ooint_{\rm FS}{\bf F(k)}\cdot d\bf S,
\end{equation}
  where FS is the curved surface surrounding the Weyl point.
  By evaluating $C_{\rm FS}$, we find that $K_+$ and $K_-$  are a pair of Weyl points with opposite charge $+1$ and $-1$, denoted by the open and solid circles respectively. This means that this 3D circuit system hosts four Weyl points that reside at the same admittance and is thus a Weyl semimetal.

\section{III. The 3D square-root HOTI}

\begin{figure}[htbp!]
 \centering
 \includegraphics[width=0.96\textwidth]{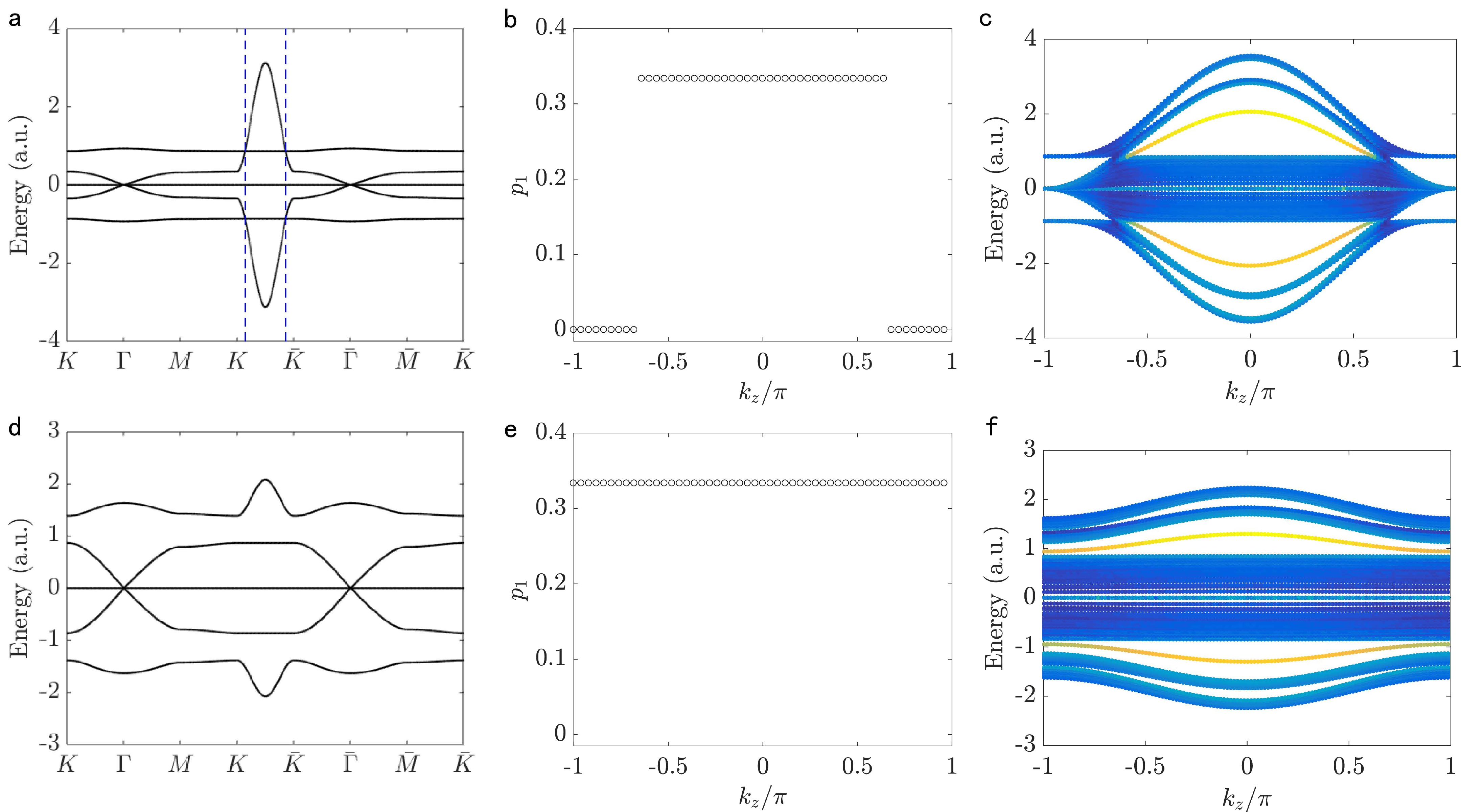}\\
 \caption{The parameters in \textbf{a}, \textbf{b}, and \textbf{c} were chosen as with $t_a=0.5$, $t_b=1$, and $t_z=0.1$, corresponding to SHOWS. As a comparison, the parameters in \textbf{d}, \textbf{e} and \textbf{f} were chosen as with $t_a=0.5$, $t_b=1$, and $t_z=0.5$, corresponding to square-root HOTI. \textbf{a}\textbf{d} Bulk band structures. \textbf{b}\textbf{e} Bulk polarization $p_1$ as a function of $k_z$, with the subscript $1$ indicating the 1th band.  \textbf{c}\textbf{f} The projected dispersion of a triangular prism, i.e., admittance along the $k_z$ direction. The yellow line indicates the hinge state dispersion. }\label{KMZ}
\end{figure}
The non-zero bulk polarization (in Fig. \ref{KMZ}\textbf{b}) gives rise to the hinge states in a triangular prism sample with the dispersion connecting the projections of the Weyl points along the $k_z$ direction, as shown by the hinge state distribution in Fig. \ref{KMZ}\textbf{c}.
It is worth mentioning that a 3D square-root HOTI can also emerge in our system for other parameters (see Figs. \ref{KMZ}\textbf{d}-\ref{KMZ}\textbf{f}). Comparing the bulk band structures in Fig. \ref{KMZ}\textbf{a} and Fig. \ref{KMZ}\textbf{d}), one can see that the band gap of high-order topological insulators always exists  from $K$ to $\bar{K}$. In this region, the bulk polarization is always non-zero in Fig. \ref{KMZ}\textbf{b}, but not the case for Fig. \ref{KMZ}\textbf{e}.

\section{IV. Calculations of the Fermi arcs}
 \begin{figure}[htbp!]
 \centering
 \includegraphics[width=0.96\textwidth]{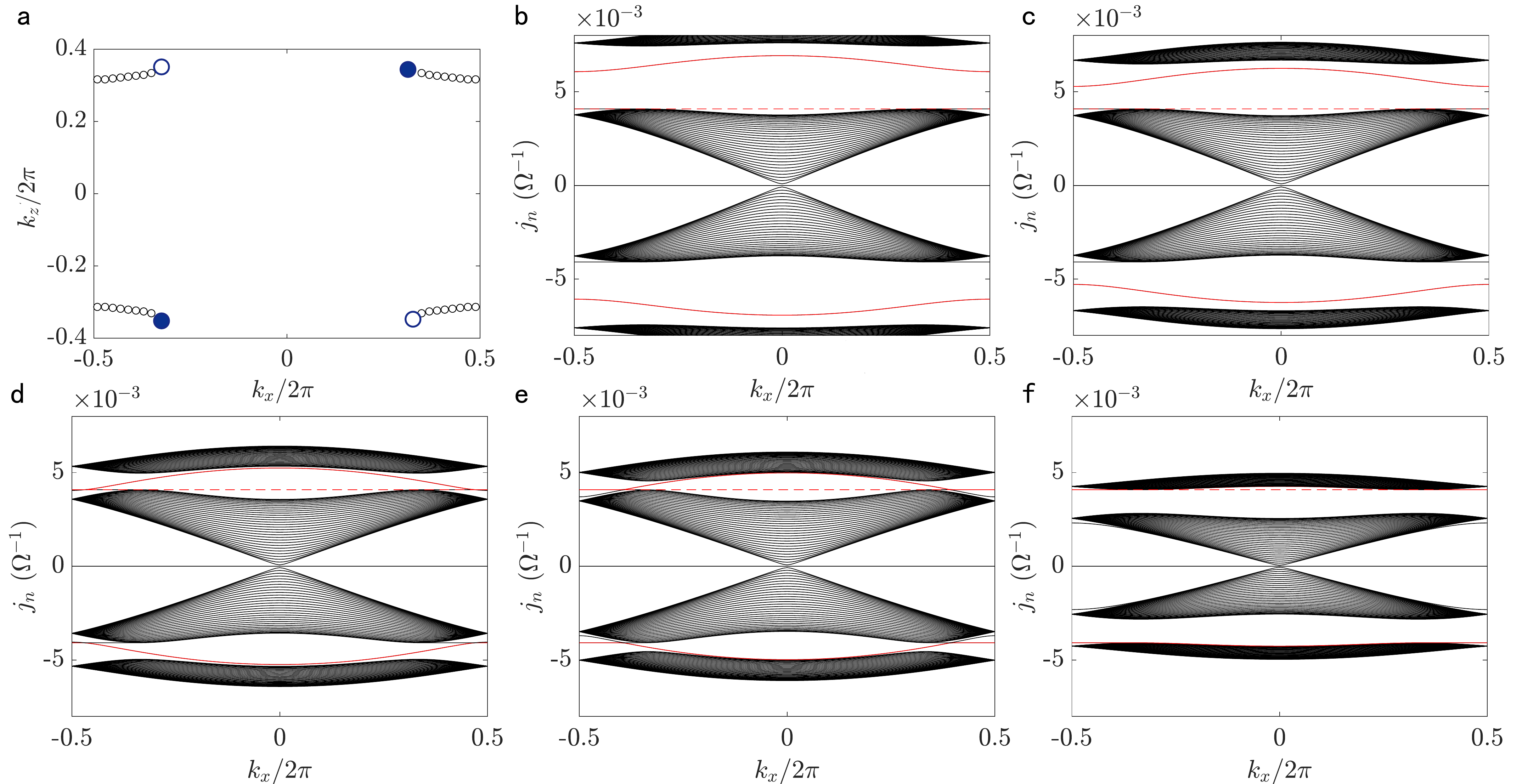}\\
 \caption{Fermi arc and surface state dispersions. \textbf{a} The contour of the surface states at the admittance of the Weyl points ($j_n=0.004082$ $\Omega^{-1}$). The open and solid circles denote the Weyl points with opposite topological charges. \textbf{b}-\textbf{f} The surface state dispersions along the $k_x$ direction for different $k_z$ (1.66, 1.78, 1.975, 2.03, and 2.283). The solid red line denotes the surface state dispersion and the dashed line shows the position of $j_n=0.004082$ $\Omega^{-1}$. }\label{Fermi_arcs}
\end{figure}
The Fermi arc is the equi-energy contour of the surface states at a fixed $j_n=0.004082$ $\Omega^{-1}$. Figure \ref{Fermi_arcs}\textbf{a} shows the Fermi arcs with the same energy of Weyl points. Because all the four Weyl points are at the same energy, the Fermi arcs connect two Weyl points with opposite charges. These surface states are clearly gapped, as shown in Figs, \ref{Fermi_arcs}\textbf{b}-\ref{Fermi_arcs}\textbf{f}.

\section{V. Mapping from Kirchhoff's law to Schr{\"o}dinger equation} \label{V}
We derive the relation between Kirchhoff's laws and Schroedinger equation, which enables us to calculate the frequency spectrum.

In electric circuits, the equation of motion is given by
\begin{equation}
\frac{d{\textbf{I}}(t)}{dt}=C\frac{d^2\textbf{V}(t)}{dt^2}+L\textbf{V}(t),
\end{equation}
where \textbf{V} is the $N$-component voltage measured at each node against the ground and \textbf{I} is the $N$-component input current at each node.

The homogeneous equations of motion ($\textbf{I}=0$) can be rewritten as $2N$ differential equations of first order \cite{Hofmann2019}:
\begin{equation}
-i\frac{d}{dt}\psi(t)=\mathcal{H}_S\psi(t),
\end{equation}
with $\psi=(\dot{\textbf{V}}(t),\textbf{V}(t))^T$ and the Hamiltonian block matrix being
$\mathcal{H}_S=i\left(
        \begin{array}{cc}
         0 & C^{-1}L \\
         -\textbf{1} & 0 \\
        \end{array}
       \right).$
By diagonalizing $\mathcal{H}_S$, we can obtain the frequency dispersion $\omega(k_x)$.